\documentclass[12pt,twoside,fleqn]{article}

\usepackage{a4wide,cite}
\usepackage{amsmath,amssymb}
\usepackage{epsfig,rotating}

\usepackage[dvips]{pstcol}

\usepackage{axodraw}

\numberwithin{equation}{section}
\allowdisplaybreaks[4]

\newcommand{\MS}{\ensuremath{\overline{\text{MS}}}}

\begin{document}

\begin{titlepage}

\vspace*{1cm}

\centerline{\Large\bf Charm quark mass dependence of the electromagnetic}
\vspace{2mm}
\centerline{\Large\bf\boldmath dipole operator contribution to $\bar B\to X_s\gamma$ at $O(\alpha_s^2)$}
\vskip 2.5cm

\begin{center}
  {\bf H.~M.~Asatrian$^a$, T.~Ewerth$^b$, H. Gabrielyan$^a$, and C.~Greub$^b$}\\[2mm]
  {$^a$\sl Yerevan Physics Institute, 375036 Yerevan, Armenia}\\[1mm]
  {$^b$\sl Institute for Theoretical Physics, Univ. Berne, CH-3012 Berne, Switzerland}
\end{center}
\medskip

\vskip 2cm

\begin{abstract}
\noindent We extend existing calculations of the electromagnetic dipole operator contribution
to the total decay rate and the photon energy spectrum of the decay
$\bar{B}\to X_s\gamma$ at $O(\alpha_s^2)$ by working out the exact dependence
on the charm quark mass.
\end{abstract}

\end{titlepage}


\section{Introduction}

As a flavor changing neutral current process the inclusive decay $\bar{B}\to X_s\gamma$ is loop-induced and therefore
highly sensitive to new degrees of freedom beyond the Standard
Model. To tap the full
potential of this decay channel in deriving constraints on the parameter
space of new physics models
both the
experiments and the Standard
Model calculations should be known as accurately as possible.

On the experimental side, the latest measurements by Belle and BABAR are reported in
\cite{Koppenburg:2004fz,Aubert:2006gg}, and the world average performed by the Heavy Flavor Averaging
Group~\cite{unknown:2006bi} for $E_\gamma > 1.6\,{\rm GeV}$ reads
\begin{equation}\label{hfag}
  \text{Br}(\bar B\to X_s\gamma)=\left(3.55\pm 0.24^{+0.09}_{-0.10}\pm 0.03\right)\times 10^{-4}\,,
\end{equation}
where the errors are statistical, systematical, due to the extrapolation to the common lower-cut in the
photon energy, and due to the $\bar{B}\to X_d\gamma$ contamination, respectively.

In order to compete with the given experimental accuracy the theoretical prediction of the $\bar{B}\to X_s\gamma$ branching
ratio has to be known at the next-to-next-to-leading order (NNLO) level. There have been great efforts of several groups
within the last few years to
achieve this goal. The three-loop dipole operator matching was found in \cite{Misiak:2004ew}, the three-loop mixing of
the four-quark operators in \cite{Gorbahn:2004my}, and the three-loop mixing of the dipole operators was calculated
in \cite{Gorbahn:2005sa}.
Furthermore, the four-loop mixing of the four-quark operators into the dipole operators was
calculated in \cite{Czakon:2006ss}.
The two-loop matrix elements of the electromagnetic dipole operator together with
the corresponding bremsstrahlung terms can be found in \cite{Blokland:2005uk,Melnikov:2005bx,Asatrian:2006ph,Asatrian:2006sm}.
The three-loop matrix elements of the four-quark operators were found in \cite{Bieri:2003ue} within the so-called
large-$\beta_0$ approximation, and a calculation that goes beyond this approximation by employing an interpolation in the
charm quark mass $m_c$ was presented in \cite{Misiak:2006ab}. The
combination of all these individual contributions culminated
in a first estimate of the $\bar{B}\to X_s\gamma$ branching ratio at $O(\alpha_s^2)$ \cite{Misiak:2006zs}. For
$E_\gamma > 1.6\,$GeV it reads
\begin{equation}\label{estimate}
  \text{Br}(\bar B\to X_s\gamma) = (3.15\pm 0.23)\times 10^{-4}\,.
\end{equation}
Here, we should mention that  
there are several perturbative and 
non-perturbative effects that have not been considered
when deriving this estimate. Some of these are already available in the
literature: the four-loop mixing of $O_1,\dots,O_6$ into $O_8$ \cite{Czakon:2006ss};
the bremsstrahlung contributions of the $(O_2,O_2)$, $(O_2,O_7)$ and $(O_7,O_8)$-interferences
at $O(\alpha_s^2\beta_0)$ \cite{Ligeti:1999ea};
those $(O_2,O_2)$ and $(O_2,O_7)$ contributions which are due
to the renormalization of $m_c$ \cite{Asatrian:2005pm} (written as expansions in $m_c/m_b$);
photon energy cut-off related effects \cite{Neubert:2004dd,Becher:2005pd,Becher:2006qw,Becher:2006pu,Andersen:2006hr};
and
estimates for the $O(\alpha_s\Lambda_\text{QCD}/m_b)$ corrections \cite{Lee:2006wn}.
Other effects are unknown at the moment, like
the complete virtual- and bremsstrahlung contributions to the $(O_7,O_8)$- and 
$(O_8,O_8)$-interferences  
at $O(\alpha_s^2)$ \footnote{ In (\ref{estimate}) the 
virtual corrections to the $(O_7,O_8)$-interference at 
$O(\alpha_s^2 \beta_0)$ were taken into account. };
 the exact $m_c$-dependence of various matrix elements beyond the large
 $\beta_0$-approximation, in order to improve (or even remove) the uncertainty
due to the interpolation in $m_c$ \cite{Misiak:2006ab}; and
the emission process of photons from four-quark operators at
 tree-level. The individual contributions listed above are all
expected to remain within the uncertainty given in 
(\ref{estimate}), nevertheless they should be taken into account
in  future updates.

In the present paper we extend the calculations of the
$(O_7,O_7)$-interference contribution
 performed in 
\cite{Blokland:2005uk,Melnikov:2005bx,
Asatrian:2006ph,Asatrian:2006sm} to include the charm quark
mass at its physical value.
Since the results
given there were presented for $N_H$ heavy quarks with masses equal to $m_b$ and $N_L$
light quarks with masses equal to zero, 
exact results for the $(O_7,O_7)$ contribution are only available
for two extreme cases. Either we can set $N_H=2$ and $N_L=3$, that is the
charm quark mass is equal to
$m_b$, or we can set $N_H=1$ and $N_L=4$, that is the charm quark is
considered to be massless.
The $N_H$ heavy quarks are of course kinematically not allowed to appear in the final state, and
hence the first possibility corresponds qualitatively to the experimental situation since events with
charmed hadrons are not included on the experimental side. However, since $m_c\approx m_b/4$ in
reality, one could choose as well the second possibility, but in this case the
inclusion of contributions from the $c\bar c$ production is required on the theoretical side in order to
get rid of infrared divergences. As argued in \cite{Asatrian:2006ph} we expect the true result
to lie somewhere in between. To check this statement, 
we calculate the exact charm quark mass dependence 
of the $(O_7,O_7)$-interference contribution to the photon energy spectrum
$d\Gamma(b\to X^{\rm partonic}_s\gamma)/dE_\gamma$ and the total decay width $\Gamma(b\to X^{\rm partonic}_s\gamma)$,
excluding charm quarks in the final state. 
The impact of the exact $m_c$-dependence on the branching ratio
will be taken into account together with other new contributions in a forthcoming analysis.

The organization of this paper is as follows. In section 2 we describe briefly the calculation
of the relevant Feynman diagrams and present our final results for the photon energy spectrum
and the total decay width. Furthermore, we comment on the numerical importance of our
result on the branching ratio $\text{Br}(\bar B\to X_s\gamma)$ in this section.


\section{Results for the charm quark contribution}

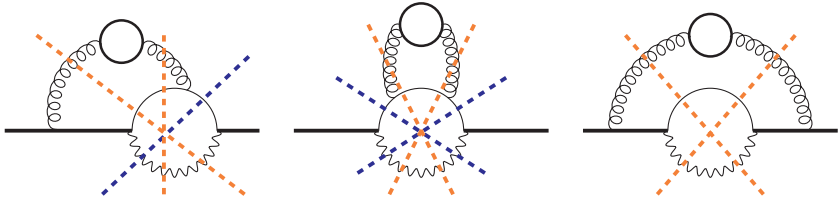
\begin{figure}[t]
  \vspace*{2cm}
  \begin{center}
    \begin{tabular}{ccccccc}
      \begin{picture}(0,0)(0,0)
        \SetScale{.8}
        \SetWidth{1.2}
        \CArc(-5,42)(10,0,360)
        \SetWidth{1.8}
        \Line(-60,0)(0,0)
        \Line(60,0)(40,0)
        \SetWidth{.5}
        \CArc(20,0)(20,0,180)
        \PhotonArc(20,0)(20,-180,0){2}{12.5}
        \GlueArc(-5,10)(32,16.5,72){3}{5}
        \GlueArc(-5,10)(32,107,197){3}{8}
        \SetWidth{1.8}
        \SetColor{Blue}
        \DashLine(-14,-30)(55,35){3}
        \SetColor{Orange}
        \DashLine(15,-30)(15,45){3}
        \DashLine(-45,45)(53,-30){3}
      \end{picture}
      & \hspace{3cm} &
      \begin{picture}(0,0)(0,0)
        \SetScale{.8}
        \SetWidth{1.2}
        \CArc(0,50)(10,0,360)
        \SetWidth{1.8}
        \Line(-60,0)(-20,0)
        \Line(60,0)(20,0)
        \SetWidth{.5}
        \CArc(0,0)(20,0,180)
        \PhotonArc(0,0)(20,-180,0){2}{12.5}
        \GlueArc(25,30)(40,150,201){3}{7}
        \GlueArc(-25,30)(40,-21,30){3}{7}
        \SetWidth{1.8}
        \SetColor{Blue}
        \DashLine(-30,-20)(40,25){3}
        \DashLine( 30,-20)(-40,25){3}
        \SetColor{Orange}
        \DashLine(-15,-30)(25,50){3}
        \DashLine(15,-30)(-25,50){3}
      \end{picture}
      & \hspace{3cm} &
      \begin{picture}(0,0)(0,0)
        \SetScale{.8}
        \SetWidth{1.2}
        \CArc(0,45)(10,0,360)
        \SetWidth{1.8}
        \Line(-60,0)(-20,0)
        \Line(60,0)(20,0)
        \SetWidth{.5}
        \CArc(0,0)(20,0,180)
        \PhotonArc(0,0)(20,-180,0){2}{12.5}
        \GlueArc(0,0)(45,0,78){3}{11}
        \GlueArc(0,0)(45,102,180){3}{11}
        \SetWidth{1.8}
        \SetColor{Orange}
        \DashLine(-25,-30)(40,45){3}
        \DashLine(25,-30)(-40,45){3}
      \end{picture}\\
      & & & & & & \\[2mm]
    \end{tabular}
  \end{center}
  \caption{\sl 2- and 3-particle-cuts of the irreducible $b$ quark selfenergy diagrams with a massive charm quark bubble
    contributing to the $b\to s\gamma$ (blue dashed line) and $b\to s\gamma g$ (orange dashed lines) transitions at
    $O(\alpha_s^2)$. Thick
    lines denote $b$ quarks, thin lines $s$ quarks, wiggly lines photons and curly lines gluons. Left-right reflected
    diagrams are not shown.}\label{charmloop}
\end{figure}

Within the low-energy effective theory the partonic $b\to X_s\gamma$ decay rate can be written as
\begin{equation}\label{decay_rate}
\Gamma(b\to X_s^{\rm parton}\gamma)_{E_\gamma>E_0} = \frac{G_F^2\alpha_{\rm
    em}\overline{m}_b^2(\mu)m_b^3}{32\pi^4}\,|V_{tb}^{}V_{ts}^*|^2\,\sum_{i,j}C_i^{\rm eff}(\mu)\,C_j^{\rm
  eff}(\mu)\,G_{ij}(E_0,\mu)\,,
\end{equation}
where $m_b$ and $\overline{m}_b(\mu)$ denote the pole and the running $\MS$ mass of the
$b$ quark, respectively, $C_i^{\rm eff}(\mu)$
the effective Wilson coefficients at the low-energy scale, and $E_0$ the
energy cut in the photon spectrum.
As already anticipated in the introduction, we will focus on the function $G_{77}(E_0,\mu)$
corresponding to the self-interference of the electromagnetic dipole operator
\begin{equation}
O_7 = \frac{e}{16\pi^2}\,\overline{m}_b(\mu)\left(\bar s\sigma^{\mu\nu}P_Rb\right)F_{\mu\nu}\,.
\end{equation}
More precisely, we extend the calculations performed in \cite{Blokland:2005uk,Melnikov:2005bx,
Asatrian:2006ph,Asatrian:2006sm} to include the effects of a massive charm quark. To
this end we calculate the cuts of the $b$ quark self-energies displayed in figure \ref{charmloop}
with two or three particles in the intermediate state. We do not have to calculate cuts with four
particles in the intermediate state since such cuts would run through the
charm quark bubble and events involving charmed hadrons in the final state
are not included on the experimental side.

The reduction of the 2-particle-cut of the first Feynman diagram
visualized in figure \ref{charmloop} to a set of a few so-called
master integrals is done by means of the systematic Laporta algorithm \cite{Laporta:2001dd}
based on the integration-by-part technique first proposed in \cite{Tkachov:1981wb,Chetyrkin:1981qh} \footnote{For an
automatized implementation of this algorithm see, e.g., \cite{Anastasiou:2004vj}.}.
Since the reduction procedure applied to the present problem has already been described in great detail
in \cite{Asatrian:2006ph,Asatrian:2006sm},
we refrain from repeating it here once more. We just remark that after the reduction there remain
only three two-loop integrals which do not factorize into a product of one-loop integrals.
To solve these integrals, we first introduce Feynman parameters in the standard way and perform
the loop-integrations. At this level, the Feynman parameter integrals contain denominators of the form
\begin{equation}
  \frac{1}{(m_c^2P_1+m_b^2P_2)^\alpha}\,,
\end{equation}
where $P_1$ and $P_2$ are polynomials in the Feynman parameters.
Next, applying the Mellin-Barnes representation \cite{Boos:1990rg}
\begin{equation}
  \frac{1}{(x+y)^\alpha} = \frac{1}{\Gamma(\alpha)}\int\limits_C\!\frac{ds}{2\pi i}
  \frac{x^s}{y^{\alpha+s}}\,\Gamma(-s)\Gamma(\alpha+s)\,,\qquad x=m_c^2P_1,\quad y=m_b^2P_2\,,
\end{equation}
where the integration contour $C$ runs from $-i\infty$ to $+i\infty$ such that it separates
the poles generated by the two $\Gamma$ functions, proves very useful because then the integration
over the Feynman parameters becomes trivial. (For explicite examples of this method, see e.g.,
\cite{Greub:1996tg}). Finally, we close the integration contour $C$ by a
half-circle with infinite radius at either
side and sum up the enclosed residues. In this way we managed to obtain solutions for the three
non-trivial two-loop integrals valid for arbitrary values of $m_c$.

The remaining 2- and 3-particle-cuts which all contain cut-diagrams with
external wave function corrections are taken into account by
proper insertions of $s$ quark and gluon wave function renormalization
constants into the corresponding 2- and
3-particle-cuts of irreducible two-loop $b$ quark self-energy diagrams. A collection of the relevant
renormalization constants necessary to render our amplitudes finite can be found in appendix \ref{app-a}.

Having briefly described the technical details of our calculation, we turn our attention to
the results for the function $G_{77}(E_0,\mu)$ which we write as an integral over the (rescaled) photon energy spectrum:
\begin{equation}\label{spectrum_def}
  G_{77}(E_0,\mu) = \int_{z_0}^{1}\!dz\,\frac{dG_{77}(z,\mu)}{dz}\,,
  \qquad z=\frac{2E_\gamma}{m_b}\,,\qquad z_0=\frac{2E_0}{m_b}\,.
\end{equation}
In NNLO approximation the photon energy spectrum can be decomposed as follows,\footnote{In the notation of
  \cite{Asatrian:2006sm} we have $\widehat H^{(1)}=4\,H^{(1)}$ and $\widehat H^{(2,i)} = 16\,H^{(2,i)}$
  for $i=\mbox{a},\mbox{na},\mbox{NL},\mbox{NH}$.}
 \begin{align}\label{spectrum_exp}
  \frac{d G_{77}(z,\mu)}{dz} &= \delta(1-z) + \frac{\alpha_s(\mu)}{4\pi}\,C_F \widehat H^{(1)}(z,\mu) +
  \left(\frac{\alpha_s(\mu)}{4\pi}\right)^2 C_F \widehat H^{(2)}(z,\mu) + O(\alpha_s^3)\,,
\end{align}
where
\begin{align}\label{colorf}
  \widehat H^{(2)}(z,\mu) &= C_F \widehat H^{(2,\mbox{{\footnotesize a}})}(z,\mu) +
  C_A \widehat H^{(2,\mbox{{\footnotesize na}})}(z,\mu)\nonumber\\[2mm]
  &\qquad+ T_R N_L \widehat H^{(2,\mbox{{\tiny NL}})}(z,\mu) + T_R N_H \widehat H^{(2,\mbox{{\tiny NH}})}(z,\mu) +
    T_R N_V \widehat H^{(2,\mbox{{\tiny NV}})}(z,\mu)\,.
\end{align}
Here, $N_L$, $N_H$ and $N_V$ denote the number of light ($m_q=0$), heavy ($m_q=m_b$), and virtual ($m_q=m_c$) quark flavors,
that is the total number of quark flavors is $N_F=N_L+N_H+N_V$. Furthermore, $\alpha_s(\mu)$ is the running coupling
constant in the $\MS$ scheme, and the numerical values of the color factors are given by $C_F=4/3$, $C_A=3$, and $T_R=1/2$.

The functions $\widehat H^{(1)}$ and $\widehat H^{(2,i)}$ ($i=\mbox{a},\mbox{na},\mbox{NL},\mbox{NH}$) appearing in
(\ref{spectrum_exp}) and (\ref{colorf}) receive contributions from the $b\to s\gamma$, $b\to s\gamma g$,
$b\to s\gamma gg$ and $b\to s\gamma q\bar q$ ($q\in\{u,d,s\}$, $m_q=0$) transitions.
They can be found in \cite{Asatrian:2006sm}.
On the other hand, the function
$\widehat H^{(2,\mbox{{\tiny NV}})}$ is completely new. It originates from the $b\to s\gamma$
and $b\to s\gamma g$ transitions with massive charm quark bubbles, see figure \ref{charmloop}.
Our result for this function reads
\begin{align}\label{h2nv}
  \widehat H^{(2,\mbox{{\tiny NV}})}(z,\mu) &= \bigg(\frac{124}{27}\,\pi^2 +
    \frac{32}{9}\left(7+\pi^2\right)\,L_\mu + \frac{16}{3}\,L_\mu^2 +
    \left(\frac{614}{27}-\frac{8}{9}\,\pi^2\right)\ln\rho + f(\rho)\bigg)\delta(1-z)\nonumber\\[2mm]
  &\quad+ \frac{16}{3}\left(2\,L_\mu-\ln\rho\right)\left[\frac{\ln(1-z)}{1-z}\right]_+ +
    \frac{28}{3}\left(2\,L_\mu-\ln\rho\right)\left[\frac{1}{1-z}\right]_+\nonumber\\[1.5mm]
  &\quad- \frac{8}{3}\left(7+z-2\,z^2-2\left(1+z\right)\ln (1-z)\right)\,L_\mu\nonumber\\[2mm]
  &\quad+ \frac{4}{3}\left(7+z-2\,z^2\right)\ln\rho - \frac{8}{3}\left(1+z\right)\ln(1-z)\ln\rho\,,
\end{align}
where
\begin{align}
  f(\rho) &= -\frac{\pi^2}{9}\left(162\sqrt{\rho}+70\,\rho^{3/2}-36\,\rho^2\right)+\frac{32}{9}\,\rho\ln\rho +
    \frac{2}{9}\left(25+27\,\rho^2\right)\ln^2\rho+\frac{4}{9}\ln^3\rho\nonumber\\[2mm]
  &\quad- \frac{4}{9}\left(31+27\,\rho^2\right)\ln(1-\rho)\ln\rho -
    \frac{4}{9}\sqrt{\rho}\left(81+35\,\rho\right)\text{artanh}(\sqrt{\rho})\,\ln\rho\nonumber\\[2mm]
  &\quad- \frac{2}{9}\left(62+81\sqrt{\rho}+35\,\rho^{3/2}+54\,\rho^2\right)\text{Li}_2(\rho) +
    \frac{8}{3}\ln\rho\,\text{Li}_2(\rho)\nonumber\\[2mm]
  &\quad+ \frac{8}{9}\sqrt{\rho}\left(81+35\,\rho\right)\text{Li}_2(\sqrt{\rho}) - \frac{16}{3}\,\text{Li}_3(\rho) +
    \frac{5578}{81}+\frac{172}{9}\,\rho\,.
\end{align}
Here,
\begin{equation}
  \rho = \frac{m_c^2}{m_b^2}\,,\qquad L_\mu=\ln\left(\frac{\mu}{m_b}\right)\,,
\end{equation}
$\mbox{Li}_3(z) = {\int}_{\!0}^z\,{\rm d}x\,\mbox{Li}_2(x)/x$, and $[\,\dots]_+$ are plus-distributions defined in the
standard way. Note that our result (\ref{h2nv}) holds for $\rho\in(0,\infty)$. The individual 2- and 3-particle-cuts
contributing to $\widehat H^{(2,\mbox{{\tiny NV}})}$ can
be found in appendix \ref{app-b}.

In the remainder of this section we comment on the numerical importance of the new contribution given
in (\ref{h2nv}). To this end we consider the function
\begin{align}\label{g770}
  G_{77}(0,\mu) &= 1 + \frac{\alpha_s(\mu)}{4\pi}\left\{\frac{64}{9}-\frac{16}{9}\pi^2 -
    \frac{16}{3} L_\mu \right\}\nonumber\\[2mm]
  &\qquad+ \left(\frac{\alpha_s(\mu)}{4\pi}\right)^2 \Big\{\left( 3.55556 N_F - 44.4446 \right) L_\mu^2 +
    \left(21.6168 N_F - 334.803 \right) L_\mu \nonumber\\[1mm]
  &\hspace{3.7cm}+ 37.8172 N_L +h(\rho) N_V - 2.16077 N_H - 519.250 \Big\}\,,
\end{align}
which follows from (\ref{spectrum_def}) when setting $E_0=0$ and performing the integration over $z$. The function
$h(\rho)$ which incorporates the $m_c$-dependence is given by
\begin{align}
  h(\rho) &= \frac{248}{81}\,\pi^2+\left(\frac{1972}{81}-\frac{16}{27}\,\pi^2\right)\ln\rho + \frac{2}{3}\,f(\rho)\,.
\end{align}
Equation (\ref{g770}) generalizes our result presented in \cite{Asatrian:2006ph} to include $N_V$ massive charm quarks.

Denoting the coefficient of $(\alpha_s(\mu)/(4\pi))^2$ in (\ref{g770}) by $X_2(N_H,N_L,N_V)$, we find (for $\mu=m_b$ and
$m_c/m_b=0.26$)
\begin{equation}
  X_2(1,4,0)=-370.142\,,\qquad X_2(1,3,1)=-386.638\,,\qquad X_2(2,3,0)=-410.120\,,
\end{equation}
that is the result with the physical charm quark mass is almost equal to the average of the two
approximations. From these numbers we conclude that the branching ratio $\text{Br}(\bar B\to X_s\gamma)$ will
stay within the errors quoted in (\ref{estimate}) when implementing the exact $m_c$-dependence of the
$(O_7,O_7)$-contribution. Therefore we do not repeat the interpolation performed in \cite{Misiak:2006ab} which would be
necessary to properly implement our new result since it also
estimates the $m_c$-dependence of the $(O_7,O_7)$-contribution.
We postpone this until more progress towards an improved evaluation of the branching ratio at NNLO has been achieved.

Before closing this section we mention that a normalization with $m_b^5$ rather than $\overline{m}_b^2(\mu)m_b^3$
is sometimes used in the definition of (\ref{decay_rate}). The relation between the
$\MS$ mass $\overline{m}_b(\mu)$ and the on-shell mass $m_b$ (including the exact dependence on $\rho$) necessary to
convert between both normalizations can be found in appendix A.


\section*{\normalsize Acknowledgements}
\vspace*{-2mm}
We would like to thank M. Misiak and M. Steinhauser for helpful discussions.
H.~M.~A. is partially supported by the ANSEF N 05-PS-hepth-0825-338 program.
T.~E. and C.~G. are supported by the Swiss National Foundation as well as EC-Contract
MRTN-CT-2006-035482 (FLAVIAnet).


\appendix
\section{Renormalization constants}\label{app-a}

The renormalization schemes applied in this work are exactly the same as
used in \cite{Asatrian:2006ph}, and the relevant renormalization
constants for $N_V=0$ can be found in appendix A of that reference. In
order to obtain the renormalization constants for $N_V\not=0$
form those given in \cite{Asatrian:2006ph} one has to proceed as follows:
(i) set $N_F$, the total number of quarks appearing there, equal to $N_H+N_L+N_V$;
(ii) add additional contributions $\delta Z_3^{\rm OS}$, $\delta Z_{2s}^{\rm OS}$ and
$\delta Z_{2b}^{\rm OS}$ to the gluon, $s$ quark and $b$ quark wave function renormalization constants,
respectively. These additional contributions are proportional to $N_V$, and
their explicit expressions read
\begin{align}
\label{appA1}
  \delta Z_3^{\rm OS} &= -\frac{4}{3}\,T_RN_V\,\Gamma(\epsilon)
    \,e^{\gamma\epsilon}\left(\frac{\mu}{m_b}\right)^{2\epsilon}\!
   \rho^{-\epsilon}\,\frac{\alpha_s(\mu)}{4\pi}+ O(\alpha_s^2)\,,\nonumber\\[4mm]
  \delta Z_{2s}^{\rm OS} &= C_FT_RN_V\,
    \frac{2\epsilon\,(1+\epsilon)(3-2\epsilon)\Gamma(\epsilon)^2\,e^{2\gamma\epsilon}}
    {(1-\epsilon)(2-\epsilon)(1+2\epsilon)(3+2\epsilon)}\left(\frac{\mu}{m_b}\right)^{4\epsilon}\!
   \rho^{-2\epsilon}\left(\frac{\alpha_s(\mu)}{4\pi}\right)^2+ O(\alpha_s^3)\,,\nonumber\\[4mm]
  \delta Z_{2b}^{\rm OS} &= C_FT_RN_V\Bigg\{
    \frac{1}{\epsilon}\,\left(1+8\,L_\mu-4\,\ln\rho\right) +
    \left(\frac{44}{3}-16\ln\rho\right)\,L_\mu + 24\,L_\mu^2\nonumber\\[1mm]
  &\hspace{2.5cm}+ \frac{443}{18}+28\,\rho+\frac{\pi^2}{3}\left(5-18\sqrt{\rho}-30\,\rho^{3/2}+12\,\rho^2\right) +
    \frac{8}{3}\left(2+3\,\rho\right)\ln\rho\nonumber\\[4mm]
  &\hspace{2.5cm}+ 2\left(2+3\,\rho^2\right)\ln^2\rho-4\left(1+3\,\rho^2\right)\ln(1-\rho)\ln\rho\nonumber\\[4mm]
  &\hspace{2.5cm}- 4\sqrt{\rho}\left(3+5\,\rho\right)\,\text{artanh}(\sqrt{\rho})\ln\rho +
    8\sqrt{\rho}\left(3+5\,\rho\right)\text{Li}_2(\sqrt{\rho})\nonumber\\[2mm]
  &\hspace{2.5cm}- 2\left(2+3\sqrt{\rho}+5\,\rho^{3/2}+6\,\rho^2\right)\text{Li}_2(\rho)\Bigg\}
    \left(\frac{\alpha_s(\mu)}{4\pi}\right)^2+ O(\alpha_s^3)\,,
\end{align}
with $\gamma\approx 0.5772$ being the Euler-Mascheroni constant and
$d=4-2\epsilon$ the space-time 
dimension. While the first two expressions in (\ref{appA1}) can be obtained in a
straightforward way, the calculation of the analytic expression 
for $\delta Z_{2b}^{\rm OS}$ is more
involved. We checked numerically that our result is in agreement
with the one given in \cite{Broadhurst:1991fy}.
We remark that the results in (\ref{appA1}) are independent
of the gauge parameter appearing in the gluon propagator.

Finally, we give the relation between the pole-mass $m_b$ and the $\MS$ mass $\overline{m}_b(\mu)$ up to two-loops
including the contribution of the $N_V$ quarks,
\begin{align}\label{MS-OS-shift}
  \frac{\overline{m}_b(\mu)}{m_b} &= 1 - C_F\,\left(4+6\,L_\mu\right)\,\frac{\alpha_s(\mu)}{4\pi}\nonumber\\[1mm]
  &\hspace{-.5cm}+ C_F\Bigg\{C_F\left(\frac{7}{8}+8\pi^2\ln 2-5\pi^2-12\,\zeta_3+21\,L_\mu+18\,L_\mu^2\right)\nonumber \\[1mm]
  &\hspace{.9cm}- C_A\left(\frac{1111}{24}+4\pi^2\ln 2-\frac{4}{3}\pi^2-6\,\zeta_3 +
    \frac{185}{3}\,L_\mu+22\,L_\mu^2\right)\nonumber \\[1mm]
  &\hspace{.9cm}+ T_RN_H\left(\frac{143}{6}-\frac{8}{3}\pi^2+\frac{52}{3}\,L_\mu+8\,L_\mu^2\right)\nonumber\\[1mm]
  &\hspace{.9cm}+ T_RN_L\left(\frac{71}{6}+\frac{4}{3}\pi^2+\frac{52}{3}\,L_\mu+8\,L_\mu^2\right)\nonumber \\[1mm]
  &\hspace{.9cm}+ T_RN_V\bigg(\frac{52}{3}\,L_\mu+8\,L_\mu^2+\frac{71}{6} + 12\,\rho +
    \frac{4}{3}\,\pi^2\left(1-3\sqrt{\rho}-3\,\rho^{3/2}+\rho^2\right)\nonumber \\[2mm]
  &\hspace{3cm}+ 4\,\rho\ln\rho + 2\,\rho^2\ln^2\rho - 4\left(1+\rho^2\right)\ln(1-\rho)\ln\rho\nonumber \\[4mm]
  &\hspace{3cm}- 8\sqrt{\rho}\left(1+\rho\right)\text{artanh}(\sqrt{\rho})\ln\rho
    + 16\sqrt{\rho}\left(1+\rho\right)\text{Li}_2(\sqrt{\rho})\nonumber \\[2mm]
  &\hspace{3cm}- 4\left(1+\sqrt{\rho}+\rho^{3/2}+\rho^2\right)\text{Li}_2(\rho)\bigg)\Bigg\}
    \left(\frac{\alpha_s(\mu)}{4\pi}\right)^2 + O(\alpha_s^3)\,,
\end{align}
where $\zeta_3$ is the Riemann zeta-function. We recalculated the term
proportional 
to $N_V$ and our finding agrees with
\cite{Gray:1990yh} (bearing in mind that a factor of $4/3$ is missing in front 
of the function $\Delta(M_i/M)$ appearing in
equation (17) of that reference, as also observed in 
\cite{Chetyrkin:1999qi}). 
The remaining terms in (\ref{MS-OS-shift}) have been taken from \cite{Gray:1990yh}.


\section{\boldmath Individual cut contributions to $\widehat H^{(2,\mbox{{\footnotesize NV}})}$}\label{app-b}

The function $\widehat H^{(2,\mbox{{\tiny NV}})}$ introduced in (\ref{colorf})
receives contributions from the 2-particle-cut of the first diagram given in figure 1, as well as
contributions from 2- and 3-particle-cuts where at least one renormalization constant proportional to $N_V$
is present. Denoting these contributions by $\widehat H^{(2,\mbox{{\tiny NV}},\mbox{{\tiny bare}})}_2$,
$\widehat H^{(2,\mbox{{\tiny NV}},\mbox{{\tiny ct}})}_2$, and $\widehat H^{(2,\mbox{{\tiny NV}},\mbox{{\tiny ct}})}_3$,
respectively, we have
\begin{equation}
  \widehat H^{(2,\mbox{{\tiny NV}})}(z,\mu) = \widehat H^{(2,\mbox{{\tiny NV}},\mbox{{\tiny bare}})}_2(z,\mu) +
  \widehat H^{(2,\mbox{{\tiny NV}},\mbox{{\tiny ct}})}_2(z,\mu) +
  \widehat H^{(2,\mbox{{\tiny NV}},\mbox{{\tiny ct}})}_3(z,\mu)\,,
\end{equation}
with the individual contributions given by
\begin{align}
  \widehat H^{(2,\mbox{{\tiny NV}},\mbox{{\tiny bare}})}_2(z,\mu) &= \bigg\{\frac{8}{3}\frac{1}{\epsilon^3} +
    \left(\frac{20}{3}+16\,L_\mu-\frac{8}{3}\ln\rho\right)\frac{1}{\epsilon^2}\nonumber\\[2mm]
  &\hspace{.8cm}+ \left(\frac{188}{9}-\frac{2}{9}\,\pi^2+\left[40-16\ln\rho\right]\,L_\mu+ 48\,L_\mu^2
    - \frac{16}{3}\ln\rho+\frac{4}{3}\ln^2\rho\right)\frac{1}{\epsilon}\nonumber\\[2mm]
  &\hspace{.8cm}+ \frac{7612}{81}-\frac{80}{9}\,\rho +
    \pi^2\,\left(\frac{73}{27}-12\sqrt{\rho}+\frac{20}{9}\,\rho^{3/2}\right) - 8\,\zeta_3\nonumber\\[2mm]
  &\hspace{.8cm} +
    \left(\frac{376}{3}-\frac{4}{3}\pi^2-32\ln\rho+8\ln^2\rho\right)L_\mu + \left[120-48\ln\rho\right]L_\mu^2 +
    96\,L_\mu^3\nonumber\\[2mm]
  &\hspace{.8cm}+ \left(\frac{10}{9}\,\pi^2 + \frac{8}{27}\,\left[7-15\,\rho\right]\right)\ln\rho +
    \frac{56}{9}\ln^2\rho-\frac{88}{9}\ln\rho\ln(1-\rho)\nonumber\\[2mm]
  &\hspace{.8cm}- \frac{8}{9}{\sqrt{{\rho}}}\left(27 - 5\,\rho\right)\text{artanh}(\sqrt{\rho})\ln\rho -
    \frac{4}{9}\left(22+27\sqrt{\rho}-5\,\rho^{3/2}\right)\text{Li}_2(\rho)\nonumber\\[2mm]
  &\hspace{.8cm}+ \frac{8}{3}\,\ln\rho\,\text{Li}_2(\rho) +
    \frac{16}{9}\sqrt{\rho}\,\left(27-5\,\rho\right)\text{Li}_2(\sqrt{\rho}) -
    \frac{16}{3}\,\text{Li}_3(\rho)\bigg\}\,\delta(1-z)\,,
\end{align}
\begin{align}
  \widehat H^{(2,\mbox{{\tiny NV}},\mbox{{\tiny ct}})}_2(z,\mu) &= \bigg\{-\frac{8}{3}\frac{1}{\epsilon^3} -
    \left(5+8\,L_\mu\right)\frac{4}{3}\frac{1}{\epsilon^2} -
    \left(\frac{188}{9}-\frac{4}{9}\,\pi^2+\frac{64}{3}\,L_\mu+\frac{64}{3}\,L_\mu^2 +
      4\ln\rho\right)\frac{1}{\epsilon}\nonumber\\[2mm]
  &\hspace{.8cm}- \frac{226}{9} + 28\,\rho + \pi^2\left(\frac{8}{3}-6\sqrt{\rho}-10\,\rho^{3/2}+4\,\rho^2\right) +
    \frac{64}{9}\,\zeta_3\nonumber\\[2mm]
  &\hspace{.8cm}- \left(\frac{520}{9}-\frac{16}{9}\,\pi^2+24\ln\rho\right) L_\mu-\frac{64}{3}\,L_\mu^2 -
    \frac{256}{9}\,L_\mu^3-\left(\frac{2}{3}-8\,\rho\right)\ln\rho\nonumber\\[3mm]
  &\hspace{.8cm}+ \left(4+6\,\rho^2\right)\ln^2\rho-4\left(1+3\,\rho^2\right)\ln\rho\,\ln(1-\rho)\nonumber\\[4mm]
  &\hspace{.8cm}- 4\sqrt{\rho}\left(3+5\,\rho\right)\text{artanh}(\sqrt{\rho})\ln\rho -
    2\left(2+3\,\sqrt{\rho}+5\,\rho^{3/2}+6\,\rho^2\right)\,\text{Li}_2(\rho)\nonumber\\[3mm]
  &\hspace{.8cm}+ 8\sqrt{\rho}\left(3+5\,\rho\right)\text{Li}_2(\sqrt{\rho})\bigg\}\,\delta(1-z)\,,
\end{align}
\begin{align}
  \widehat H^{(2,\mbox{{\tiny NV}},\mbox{{\tiny ct}})}_3(z,\mu) &=
    \bigg\{\left(-2\,L_\mu+\ln\rho\right)\frac{8}{3}\frac{1}{\epsilon^2}\nonumber\\[2mm]
  &\hspace{.8cm}- \left(\frac{2}{9}\,\pi^2+\left(\frac{56}{3}-16\ln\rho\right)\,L_\mu+\frac{80}{3}\,L_\mu^2 -
    \frac{28}{3}\ln\rho+\frac{4}{3}\ln^2\rho\right)\frac{1}{\epsilon}\nonumber\\[2mm]
  &\hspace{.8cm}-\frac{7}{9}\,\pi^2-\left(\frac{128}{3}-\frac{28}{9}\,\pi^2-56\ln\rho+8\ln^2\rho\right)L_\mu -
    \left(\frac{280}{3}-48\ln\rho\right)\,L_\mu^2\nonumber\\[2mm]
  &\hspace{.8cm}-\frac{608}{9}\,L_\mu^3 +
    \left(\frac{64}{3}-2\,\pi^2\right)\ln\rho-\frac{14}{3}\ln^2\rho+\frac{4}{9}\ln^3\rho +
    \frac{8}{9}\,\zeta_3\bigg\}\,\delta(1-z)\nonumber\\[2mm]
  &\hspace{.5cm}+ \frac{16}{3}\left(2\,L_\mu-\ln\rho\right)\,\left[\frac{\ln(1-z)}{1-z}\right]_+ +
    \frac{28}{3}\left(2\,L_\mu-\ln\rho\right)\,\left[\frac{1}{1-z}\right]_+\nonumber\\[2mm]
  &\hspace{.5cm}- \frac{8}{3}\left(7+z-2\,z^2-2\left(1+z\right)\ln(1-z)\right)\,L_\mu\nonumber\\[2mm]
  &\hspace{.5cm}+ \frac{4}{3}\left(7+z-2\,z^2\right)\,\ln\rho - \frac{8}{3}\left(1+z\right)\ln(1-z)\ln\rho\,.
\end{align}
The three contributions given above are by themselves independent of the gauge parameter entering the gluon propagator.


\end{document}